\documentclass[twocolumn, prl, aps, superscriptaddress, longbibliography,showpacs,amsmath,amssymb,floatfix]{revtex4-1}
\usepackage{changes}
\usepackage{graphicx}
\usepackage{dcolumn}
\usepackage{bm}
\usepackage{graphicx}                                   
\usepackage{amssymb}

\usepackage{amsmath}
\usepackage{epsfig}
\usepackage{xcolor}
\usepackage{tabu}
\usepackage{mathtools}
\usepackage[colorlinks,linkcolor=blue,anchorcolor=blue,citecolor=blue,urlcolor=blue]{hyperref}
\usepackage{physics}
\usepackage{float}
\usepackage{diagbox}
\usepackage{inputenc}

\begin{document}

\title{Brownian motion and generalized Lifson-Jackson formula in quasi-periodic systems}
\author{Sang Yang}
\affiliation{CAS Key Laboratory of Quantum Information, University of Science and Technology of China, Hefei 230026, China}
\author{Juyuan Sun}
\affiliation{CAS Key Laboratory of Quantum Information, University of Science and Technology of China, Hefei 230026, China}
\author{Guangcan Guo}
\affiliation{CAS Key Laboratory of Quantum Information, University of Science and Technology of China, Hefei 230026, China}
\affiliation{Hefei National Laboratory, University of Science and Technology of China, Hefei 230088, China}
\affiliation{Synergetic Innovation Center of Quantum Information and Quantum Physics, University of Science and Technology of China, Hefei 230026, China}
\author{Ming Gong}
\email{gongm@ustc.edu.cn}
\affiliation{CAS Key Laboratory of Quantum Information, University of Science and Technology of China, Hefei 230026, China}
\affiliation{Hefei National Laboratory, University of Science and Technology of China, Hefei 230088, China}
\affiliation{Synergetic Innovation Center of Quantum Information and Quantum Physics, University of Science and Technology of China, Hefei 230026, China}
\date{\today }
\date{\today }
	
\begin{abstract}
Brownian motion in periodic potentials has been widely investigated in statistical physics and related 
interdisciplinary fields. In the overdamped regime, it has been well-known that the diffusion constant $D^*$ is given by the Lifson-Jackson (LJ) formula. With a tilted potential, this model can exhibit giant diffusion. In this work, we start from the basic argument that 
since any quasi-periodic potential can be approximated accurately using a periodic potential, this formula and the associated physics should also apply to the quasi-periodic potential after some proper redefinition. We derive $D^*$ from the Smoluchowski equation using the fact that its asymptotic solution is a product of a Boltzmann weight and a Gaussian envelope function. Then we analytically calculate $D^*$ in terms of Bessel functions. Finally, we study the giant diffusion with quasi-periodic potentials, generalize the corresponding formula to the condition with tilted potential under the same argument, and calculate $D^*$ analytically. This work generalizes the Brownian motion from periodic potentials to the much broader quasi-periodic potentials, which should have applications in interdisciplinary fields in physics, chemistry, engineering, and life sciences. 
\end{abstract}
\maketitle

Brownian motion, which studies the erratic random motion of microscopic particles suspended in a fluid, is one of the most profound concepts in statistical physics and related 
interdisciplinary fields \cite{Bian2016111Years, ChandrasekharReview, mel1991kramers, reimann2002brownian, 
hanggi2009artificial, hanggi1990reaction}. In free space, this dynamics has been solved by Einstein by 
establishing a direct relation between the microscopic dynamics and diffusion equation, yielding the Einstein
relation $\langle x^2\rangle  =2D_0t$, where $D_0$ is the diffusion constant, $t$ is the evolution
time and $\langle x^2\rangle$ is the  mean square displacement (MSD) averaged over all possible trajectories 
\cite{einstein1905molekularkinetischen}. 
Further, the probability distribution function (PDF) of this Brownian motion 
in a harmonic potential has been calculated based on the Fokker-Planck equation by Wang and 
Uhlenbeck \cite{WangMingChen1945Theory}, and in recent years this model has been widely studied in experiments using 
levitated particles \cite{Raynal2023Shortcuts, Gieseler2012Subkelvin, Kamba2023Revealing}, with ambition to explore 
the possible physics in the quantum regime. For the particles in the periodic potential, Lifson and Jackson (LJ) have 
derived the following formula in the overdamped regime \cite{lifson1962self, FESTA1978229,  gunther1979mobility, weaver1979effective}
\begin{equation}
	D^* = {D_0 \over \langle \exp(\beta U)\rangle\langle \exp(-\beta U)\rangle},
	\label{eq-LJ-formula}
\end{equation}
where $U(x)$ is the periodic potential, $U(x) = U(x+R)$, with $R$ being the period, $\beta =1/k_B T$ 
with $k_B$ and $T$ being the Boltzmann constant and temperature, 
respectively, and the average is taken in a full period. These results have been applied in Josephson junctions \cite{Ambegaokar1969Voltage, jack2017quantum, 
bishop1978josephson, coffey2009nonlinear,longobardi2011thermal,zwerger1987quantum}, surface atom diffusion \cite{speer2009directing,AlaNissila01052002,Lacasta2004Fromsubdiffusion,Sancho2004Diffusion}, thermal ratchet \cite{reimann2002brownian, hanggi2009artificial,Bier01111997,Renzoni01052005Cold,Julicher1997Modeling} and particle dynamics in corrugated channels \cite{Palastro2008Pulse, Yang2019Diffusion, Ghosh2012Brownian} and Fick-Jacobs diffusion in confined channels \cite{Reguera2001Kinetic,Jeon2010Fractional}, cold atoms \cite{Kindermann2017Nonergodic}, just to mention a few. 

The central idea of this work is to generalize the above LJ formula to more general aperiodic potentials. Up to date, all the numerical (based on cosine and sawtooth potentials) and theoretical analysis have shown that the LJ formula is exact for any periodic function. Our key idea is based on the following arguments.  (I) The LJ formula is exact for any periodic potential; (II) any quasi-periodic potential can be accurately approximated using some periodic potential; and (III) the MSD, $\langle x^2\rangle$, is averaged over all trajectories, thus the tiny difference between the quasi-periodic potential and the periodic potential can not be distinguished. Here (II) and (III) should be a direct consequence of (I) by logical reasoning. However, in Eq. \ref{eq-LJ-formula}, the average $\langle e^{\pm \beta U(x)}\rangle$ is taken in a full period, thus if the above three arguments hold, then the LJ formula should be redefined in accordingly for the quasi-periodic potentials. With this idea, we expect that all the intriguing results in the periodic potential models should be observed with quasi-periodic potentials. Meanwhile, by the Einstein relation, we expect the PDF to be related to the Gaussian function somehow. These are the main motivations of this manuscript. 

\begin{figure}
\centering\includegraphics[width=0.45\textwidth]{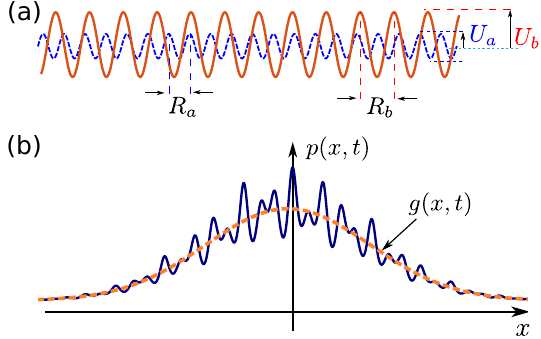}
\caption{(a) The quasi-periodic potential, with $U_a$, $U_b$ and $R_a$, $R_b$ are the potential strength period for the two components, respectively. (b) Probability $p(x, t)$ and its Gaussian envelope function $g(x, t)$ when the 
width $\sqrt{\langle x^2\rangle} \gg R_a, R_b$. }
\label{fig-fig1}
\end{figure}

This work generalizes the LJ formula to the quasi-periodic potentials, showing that its form is unchanged, yet the average should be properly defined. We then derive the LJ formula based on a new approach, showing how the Gaussian envelope function in the PDF attributes to the Einstein relation. In the quasi-periodic potential, we derive the diffusion constant with the aid of the Jacobi-Anger expansion method, yielding a compact expression of $D^*$. This result essentially comes from the ergodicity of the quasi-periodic potential during the calculation of the mean value in the LJ formula. Finally, we study the giant diffusion with quasi-periodic potential, in which $D^*$ can also be calculated using the same technique. The results in this work have greatly expanded the application range of the LJ formula and the giant diffusion, therefore 
can greatly broaden our understanding of Brownian motion in the general potential models and find intriguing applications in interdisciplinary fields \cite{Bian2016111Years, ChandrasekharReview, mel1991kramers, reimann2002brownian, 
hanggi2009artificial, hanggi1990reaction}.

{\it Physical model and diffusion constant $D^*$}. 
We start from the stochastic Langevin equation $m\ddot{x}=-\gamma \dot{x}-\frac{\text{d}U(x)}{\text{d}x}+\xi$, where 
$m$ is the mass of the particle, $\gamma$ is the friction rate, 
$U(x)$ is the periodic potential and $\xi$ is the random force. Here we are interested in the overdamped regime, in which $m\ddot{x} \simeq 0$, thus we have the following first-order Langevin equation as
\begin{equation}
\gamma \dot{x}=-\frac{\text{d}U(x)}{\text{d}x}+\xi.
\label{eqn_brown_over}
\end{equation}
with white noise defined as  
\begin{equation}
\langle \xi(t)\rangle  =0, \quad \langle \xi(t) \xi(t')\rangle = 2\gamma k_B T \delta(t-t').
\end{equation}
For the periodic potential $U(x)$ with period $R$, the average in the LJ formula is defined as 
\begin{equation}
	\langle \exp(\pm \beta U)\rangle = {1\over R} \int_0^R \exp( \pm \beta U(x))dx.
\end{equation}
This value means potential detail is not essential for $D^*$ in Brownian motion. 

In the following, we derive the above result using a different approach, from the following Smoluchowski (or  Fokker-Planck) equation
\begin{equation}
\frac{\partial}{\partial t}p(x,t)=D_0\frac{\partial }{\partial x}\Big\{e^{-\beta U(x)}\frac{\partial }{\partial x}[e^{\beta U(x)}p(x,t)]\Big\},
\label{eq-Smoluchowski}
\end{equation}
where $p(x, t)$ is the PDF and $\beta =1/k_B T$. Let us consider the overdamped regime much more carefully. We see that the requirement that $m\ddot{x} \simeq 0$ means that the particle at any position will quickly approach its 
equilibrium distribution. Furthermore, the Einstein relation $\langle x^2\rangle = 2D^*t$ in the long time limit is a 
typical results of the Gaussian distribution function. Following \cite{sivan2018probability, sivan2019diffusion, 
defaveri2023brownian}, we expect  
\begin{equation}
p(x,t)=Z e^{-\beta U(x)}g(x,t),
	\label{eq-pxt}
\end{equation}
where $Z = 1/\langle e^{-\beta U}\rangle$ is the normalization constant. In this equation, the first part will be termed as Boltzmann weight, accounting for physics in the range $|x| \ll \sqrt{4D^* t}$, and the second part is the Gaussian envelope function for physics in $|x| \sim \sqrt{4D^* t}$,
contributing to $D^*$ and the Einstein relation \cite{defaveri2023brownian}. From 
\begin{equation}
	g(x, t) = {1\over \sqrt{4\pi D^*t}} \exp(-{x^2 \over 4D^*t}),
	\label{eq-gxt}
\end{equation}
which is normalized, it yields $\langle x^2\rangle = 2D^*t$.  

\begin{figure}[htp]
\centering\includegraphics[width=0.49\textwidth]{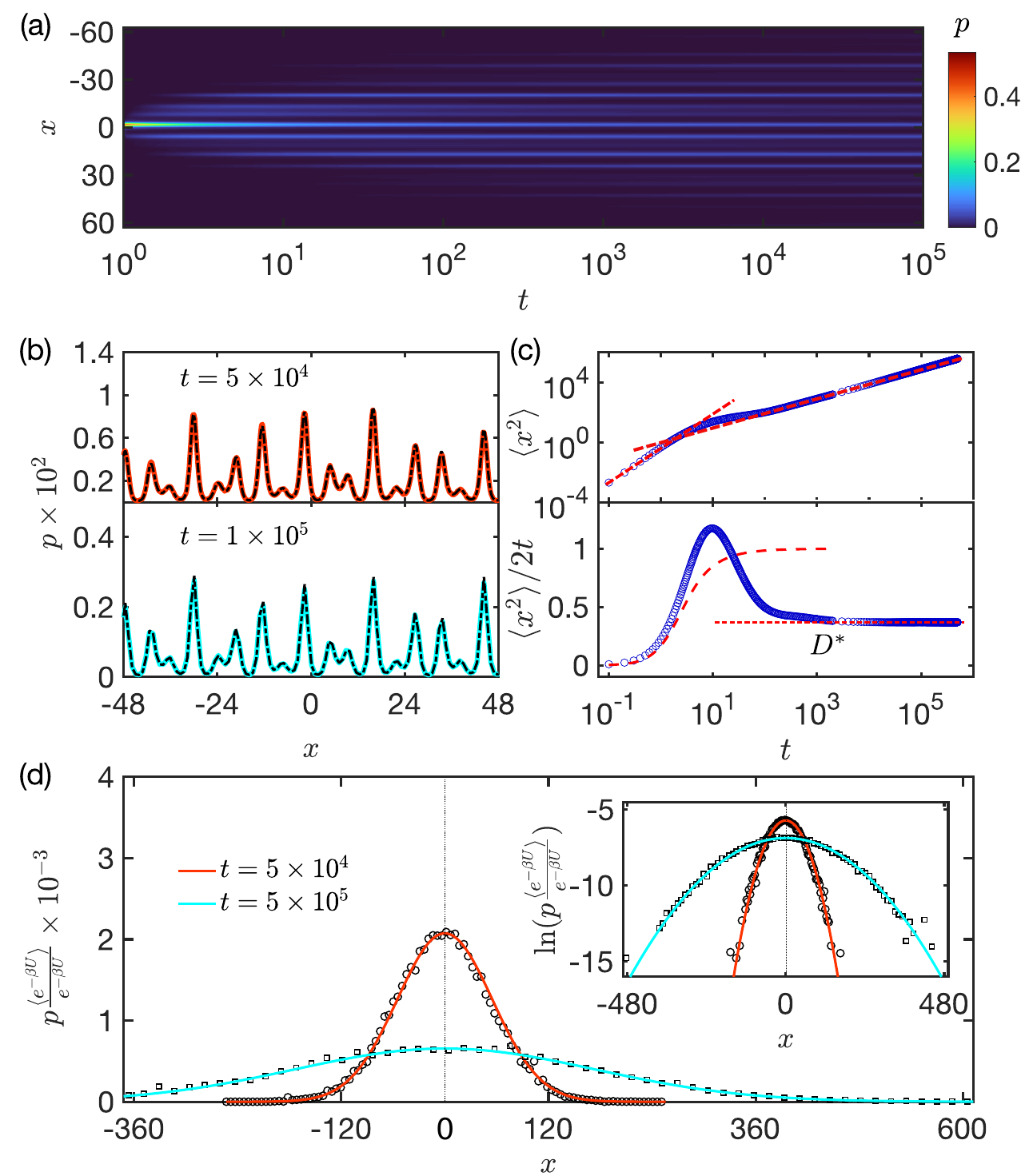}
\caption{Diffusion in one-dimensional quasi-periodical potential $U(x) = \cos(x) + \cos((\sqrt{5}-1)x/2)$. (a) Evolution of PDF $p(x,t)$. (b) PDF $p(x,t)$ at time $t=5\times 10^{4}$ and $10^{5}$. (c) Evolution of MSD $\langle x^2 \rangle$ and time-dependent diffusion constant $D^{*}(t)=\langle x^2 \rangle/2t$, with $D^* = \lim_{t\rightarrow \infty} D^*(t)$. (d) Gaussian envelope function $g(x,t)$.}
\label{fig-fig2}
\end{figure}

Next, we derive Eq. \ref{eq-gxt} and determine the associated expression of $D^*$ from the Smoluchowski equation. The key point
is that when $t$ is large enough, $g(x, t)$ can be regarded as a smooth function of $x$. For this reason 
\begin{equation}
	e^{-\beta U(x)} {\partial g \over \partial t} = D_0\frac{\partial }{\partial x}\Big\{e^{-\beta U(x)}\frac{\partial }{\partial x}
	[g(x,t)]\Big\}.
\label{eq-Smoluchowski2}
\end{equation}
Then we make an integration of both sides in a full period $R$, in which the right-hand side is in 
a total differentiation form, yielding 
\begin{equation}
	\Big\{ \int_{x}^{x+R} e^{-\beta U(z)}dz \Big\} {\partial g \over \partial t} = D_0 \Big\{ e^{-\beta U(x)} 
	\frac{\partial g(x,t)}{\partial x} \Big\}|_{x}^{x+R}.
\label{eq-Smoluchowski3}
\end{equation}
Further, since $g(x, t)$ is a smooth function of $x$ and $t$, the right hand side should be equal to 
$(\partial^2 g(x,t)/\partial x^2)R$. Finally, we move $e^{-\beta U}$ to the left-hand side and repeat the 
above procedure, we obtain the following equation 
\begin{equation}
	{\partial g \over \partial t} = D^* {\partial^2 g \over \partial x^2},
\end{equation}
which naturally yields the Einstein relation. This result may also imply that if $\langle x^2\rangle = 
2D^* t$, the PDF may related somehow to the Gaussian envelope function. With this solution and using the 
normalization condition, we can immediately determine $Z = \langle e^{-\beta U}\rangle$, which is the mean value 
in a full period. The above analysis 
represents a new angle to understand the LJ formula, which is much more straightforward as compared with the 
previous ones \cite{lifson1962self, FESTA1978229,  gunther1979mobility, weaver1979effective}.

The above derivation contains a tricky point, which is not self-evident. In the last step, we have to move $e^{-\beta U}$ to 
the left-hand side. The major reason is that Eq. \ref{eq-pxt} is only the asymptotic solution of the Smoluchowski equation. 
The exact solution at finite $t$ contains some extra terms related to $f_n(x)/t^{n}$ \cite{sivan2018probability}, 
in which when the derivative is taken 
these terms will have leading contributions. To make the right-hand side a total differentiation form, we have to move 
this term to the left-hand side. In this way, $D^*$ is unchanged upon change $U \rightarrow \pm U + \mathcal{C}$, where 
$\mathcal{C}$ is an arbitrary constant. 

{\it Generalization to quasi-periodic potential}. This new derivation of the PDF and $D^*$ will be useful for us to understand the results in the quasi-periodic potentials. The major logical reason is that the solution of the Smoluchowski equation and the associated Gaussian envelope function does not require the potential to be periodic. There are a lot of similar circumstances in scientific research that when we derive a conclusion from some specific assumptions, it is quite possible that these assumptions are not essential or even not necessary for this conclusion. We are in the same situation, yet for a long time, this point has not yet been seriously considered.

We find that for the quasi-periodic potentials, the LJ formula is still correct, upon the following definition
\begin{equation}
	\langle e^{\pm \beta U(x)}\rangle 
	= \lim_{L\rightarrow \infty} {1 \over L} \int_{x}^{x+L} e^{\pm \beta U(z)}dz.
	\label{eq-meanUxLlimit}
\end{equation}
In this way, we can still use Eq. \ref{eq-LJ-formula} to obtain $D^*$, yet the average should be carried out in 
the interval $[x, x+L]$, where $L$ should be large enough to ensure that the above mean value converges. This is
of course a direct logical consequence of the three arguments (I) - (III), which motivate this manuscript.

We verify the above results in Fig. \ref{fig-fig2} using a potential $U(x) = U_a \cos(ax) + U_b \cos(bx)$, in which $a/b$ is an irrational constant, with periods $R_a = 2\pi/a$ and $R_b =2\pi/b$ (see Fig. \ref{fig-fig1}(a)). The evolution of 
PDF $p(x, t)$ is presented in Fig. \ref{fig-fig2}(a) and Fig. \ref{fig-fig2}(b), and the MSD and the associated $D^*$
are presented in Fig. \ref{fig-fig2}(c), in which the numerical and theoretical values agree with extremely high accuracy. Especially, we find that the effect of potential will become important when $\langle x^2\rangle < \text{min}\{R_a^2, R_b^2\}$. Finally, we calculate the envelope function, determined by $p/\exp(-\beta U)$, 
demonstrating that it is a Gaussian function in Fig. \ref{fig-fig2}(d). In our simulation, we find $L \sim (10 - 20) R_a$ 
or $(10 -20)R_b$ is sufficient to obtain a accurate approximation of $\langle e^{\pm \beta U(x)}\rangle$. 

With this new definition, we can calculate $D^*$ analytically using Eq. \ref{eq-meanUxLlimit} in 
the large $L$ limit. The key point is that during the integration of $x \in [0, L]$, with $L \rightarrow \infty$, the
two potentials $U_a\cos(ax)$ and $U_b\cos(bx)$ are un-correlated in the large $x$ limit and they will uniformly traverse
the entire space, leading to $\langle e^{\pm \beta U}\rangle = \langle e^{\pm \beta U_a(x)}\rangle \times \langle e^{\pm \beta U_b(x)}\rangle$, 
therefore
\begin{equation}    D^* = {D_0 \over I_0^2(\beta U_a) I_0^2(\beta U_b)},
    \label{eq-DstarBesselI0}
\end{equation}
where $I_0$ is the modified Bessel function of the first kind. We can derive the above intuitive result using the following 
Jacobi-Anger expansion (assuming Einstein summation rule) $e^{A \cos(ax)} = I_n(A) e^{inax}$, where $I_n(A)$ is the modified Bessel function of the first kind. It yields
\begin{equation}
	{1\over L} \int_0^L e^{\beta U(x)}dx = \sum_{nm} I_n(\beta U_a) I_m(\beta U_b) {1\over L} \int_0^L e^{n a x + m b x}dx.
	\label{eq-Besselexpansion}
\end{equation}
Since $a$ and $b$ are incommensurate numbers, in the above integration only $n = m = 0$ survives, leading to 
Eq. \ref{eq-DstarBesselI0}. This diffusion constant can be obtained in much more aperiodic potentials. 
For instance, if $U(x) = \sum_i U_i \cos(k_i x + \phi_i)$, where all $k_i$ are mutual incommensurate constant, and 
$\phi_i$ are arbitrary phases, we have 
\begin{equation}
	D^* = {D_0 \over \prod_i I_0^2(\beta U_i)}.
\end{equation}
This result means that our results can be further extended to arbitrarily aperiodic yet 
bounded potentials, which represents a remarkable extension of the LJ formula.

\begin{figure}[htp]
\centering\includegraphics[width=0.49\textwidth]{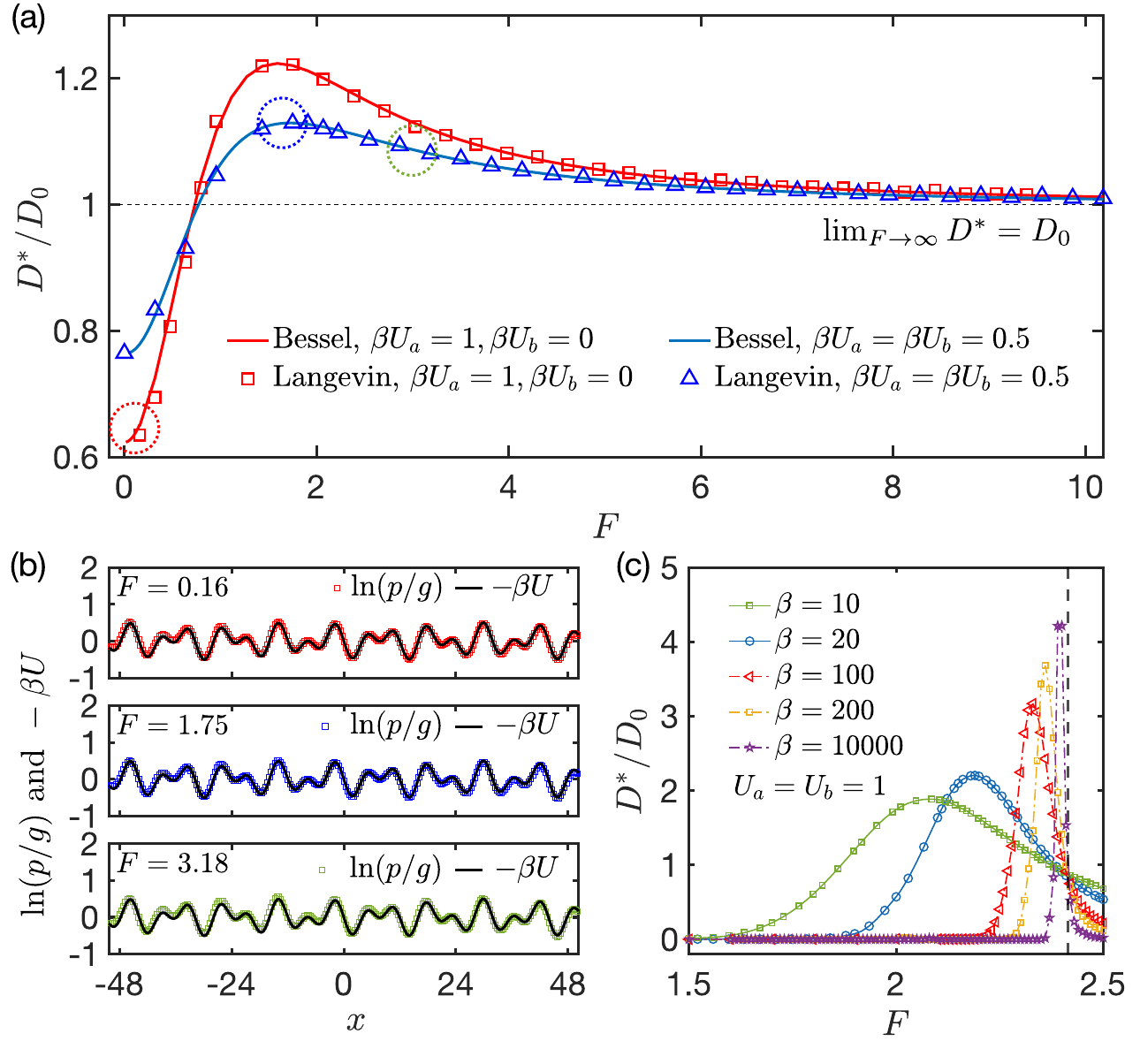}
\caption{Giant diffusion with quasi-periodic potential $U(x) = U_a \cos(x) + U_b \cos(bx)$. (a) $D^*$ in different tilde slopes $F$ and potential strengths. The symbols are from the Langevin equation and solid lines are obtained by Bessel expansion. (b) The Boltzmann weight of the wave function for three fields $F$, see dashed circles in (a). (c) Giant 
diffusion and critical $F_c$ for various temperatures. The vertical dashed line gives $F_c = U_a + \sqrt{2}U_b$ in the low-temperature limit \cite{Reimann2001giant,  Reimann2008Weak,Reimann2002diffusion}. }
	\label{fig-fig5}
\end{figure}

{\it Giant diffusion with a titled potential}: With the above results, we next turn to the physics of giant diffusion, 
which has been intensively studied in a tilted periodic potential, such as $U(x) = U_a \cos(ax) + Fx$ \cite{Lindner2016Giant, Lindner2008Critical, Spiechowicz2020Diffusion,  Dan2002Giant, Reimann2001giant,  Reimann2008Weak, Reimann2002diffusion, Iida2025Universality}, where $F$ is the field strength. 
This model can be related a particle moves on a tilted surface under the action of gravitation. In this model, it has
been found that the diffusion constant will take a maximal value at some critical strength $F_c$. However, a close 
analysis of this problem also reveals that periodicity is not essential for this giant diffusion. It only brings some 
convenience to calculate $D^*$. Thus we are in the same circumstance to investigate the giant diffusion with a quasi-periodic potential. 

Our basic reasoning is the same as arguments (I) - (III). It has been found from the first passage time  that in the presence of a tilted field and in the long time limit \cite{Reimann2001giant,  Reimann2008Weak, Reimann2002diffusion, hanggi1990reaction}
\begin{eqnarray}
	D^{\ast} = D_0 \frac{\int_{x_0}^{x_0+L} \frac{d x}{L} I_{ \pm}(x) I_{+}(x) I_{-}(x)}{\left[\int_{x_0}^{x_0+L} \frac{d x}{L} I_{ \pm}(x)\right]^3}, 
\end{eqnarray}
where 
\begin{equation}
		I_{\pm}(x)= \int_0^L dy e^{\pm \beta U(x) \mp \beta U(x-y) - \beta F y},
\end{equation}
for $F > 0$, with $Fx$ being the tilt potential. We can write the above form in the following form 
\begin{equation}
			D^{\ast} =  D_0 {\langle I_{ \pm}(x) I_{+}(x) I_{-}(x)\rangle \over (\langle I_{ \pm}(x)\rangle)^3}.
\end{equation}
This result can also be calculated using the same expansion in Eq. \ref{eq-Besselexpansion}. When $F = 0$, we have
$I_+  = e^{\beta U(x)} \langle e^{-\beta U(x)} \rangle L$, and 
$I_-  = e^{-\beta U(x)} \langle e^{\beta U(x)} \rangle L$, based on which the above formula will immediately reduced to the LJ formula in Eq. (\ref{eq-LJ-formula}). In this result, the only requirement is $\langle e^{\pm U(x)}\rangle$ to be existed, which is a rather natural conclusion for the bounded and aperiodic potentials, not necessary to be quasi-periodicity. 

We illustrate the result using a single potential $U(x) = U_a \cos(ax)$ and then give the final results in the quasi-periodic 
potentials. We have
\begin{equation}
	I_+ = e^{\beta U_a(x)} \mathcal{\bar{G}}_{na} e^{inax}, \quad 
	I_- = e^{-\beta U_a(x)} \mathcal{G}_{na} e^{inax},
	\label{eq-gsingle}
\end{equation}
where $\mathcal{G}_{na} = I_{na}/(ina + \beta F)$,  $\mathcal{\bar{G}}_{na} = I_{-na}/(ina + \beta F)$, and $I_{na} = I_n(\beta U_a)$. In this way, we have 
\begin{equation}
	D^* = D_0 {  \mathcal{\bar{G}}_{na} \mathcal{\bar{G}}_{ma} \mathcal{G}_{ka} I_{-(n+m+k)a}
	\over (\mathcal{\bar{G}}_{na} I_{-na})^3}.
	\label{eq-Dsingle}
\end{equation}
We see that using the Jacobi-Anger expansion, the field $F$ plays the role of phase matching between different 
plane waves, which is essentially for the giant diffusion. From this angle, it is the scattering between these 
different plane wave bases that contributes to the giant diffusion, thus the periodicity of the potential is not 
that crucial.

We can use the same method to obtain the $D^*$ with two incommensurate periodic potentials \cite{Lopez2020Enhanced}
\begin{equation}
U(x) = U_a \cos(ax) + U_b \cos(bx),
\end{equation}
by using  
$I_+(x) = e^{\beta U(x)} e^{i(na+n'b)x} \mathcal{\bar{G}}_{na,n'b}$, and 
$I_-(x) = e^{-\beta U(x)} e^{i(na+n'b)x} \mathcal{G}_{na,n'b}$, where $\mathcal{\bar{G}}_{na,n'b} 
=I_{-na} I_{-n'b}/(ina + in'b + \beta F)$. Thus the field $F$ plays a role in coupling different plane waves, which contribute to the giant diffusion. Compared with Eq. \ref{eq-gsingle} and Eq. \ref{eq-Dsingle}, 
we find that in the presence of two periods, only the $\mathcal{G}$ functions have to be redefined, thus
\begin{equation}
	D^* = D_0 {  \mathcal{\bar{G}}_{na,n'b} \mathcal{\bar{G}}_{ma,m'b} \mathcal{G}_{ka,k'b} I_{-(n+m+k)a} I_{- (n'+k' + b')b}
    \over (\mathcal{\bar{G}}_{na,n'b} I_{-na} I_{-n'b})^3}.
    \label{eq-Ddouble}
\end{equation}
We see that using the field $F$, some of the scattering between the plane waves will be allowed, yielding giant diffusion. A finite truncation $|n|$, $|n'| < N_c$, with $N_c \sim 3 - 5$ may be obtain a sufficiently accurate $D^*$. 

The numerical results for $D^*/D_0$ are presented in Fig. \ref{fig-fig5} as a function of $F$, yielding a peak at $F_c$. As we expected, with the Einstein relation, we believe that the PDF should be related to the Gaussian envelope function in some
way. We find that 
\begin{equation}
	p(x, t) = Z e^{-\beta U(x)} g(x-x_c, t),
\end{equation}
with $x_c = \beta D^*Ft$, we can obtain the Boltzmann weight using $p/g$, which is presented in Fig. \ref{fig-fig5}(b). This result implies that the Boltzmann weight is independent of $F$. Finally, at extremely low temperatures and when $F < F_c = U_a a + U_b b$ \cite{Reimann2001giant, Reimann2008Weak, Reimann2002diffusion}, the particles will be trapped in the local minimal, 
thus $\lim_{T\rightarrow 0 } D^* = 0$. It will become nonzero only when $F > F_c$, the directional transport is allowed. When 
$F \ll F_c$, we will find that $D^* \rightarrow D_0$, which is shown in Fig. \ref{fig-fig5}(c). This can be understood from Eq. \ref{eqn_brown_over}, in which when $F$ is dominated, we have $\gamma \dot{x} = F + \xi$, yielding $D^* = D_0$ and $x_c = Ft/\gamma$ \cite{Reimann2001giant,  Reimann2008Weak,Reimann2002diffusion}. The results in Fig. \ref{fig-fig5} agree with these analyses with extremely high accuracy.

{\it Conclusion and Discussion}: To conclude, we generalize the Brownian motion from the periodic systems to the quasi-periodic systems and find that the 
LJ formula, after some proper redefinition, can be used to describe their dynamics. This is expected 
since the quasi-periodic potential can be well approximated using some proper periodic function. This result is stimulating, because it means that in some of the general bounded aperiodic potentials when the expansion of the PDF is large enough, the details of the potentials will become unimportant, thus we will always obtain the Einstein relation. In this way, the PDF should be related to the Gaussian envelope in some form. We use the same technique to explore the giant diffusion with quasi-periodic potentials. From these results, we see that the average means that the detail of the potentials is coarse-grained, thus it is quite possible that two different potentials may have the same diffusion constant $D^*$ and Gaussian envelope, with only difference in their Boltzmann weights. Our results demonstrate the much broader applicability of the LJ formula, which should greatly enhance our understanding of the Brownian motion in much more general potentials. Our result also raises an inverse yet open problem that, if the Einstein relation $\langle x^2\rangle = 2D^*t$ holds in the long time limit, does the PDF contains a Gaussian envelope? 

Intriguing application of our results can be carried out immediately. Firstly, we can obtain the general PDF with spatially quasi-periodic noise under different interpretations \cite{Pacheco2024langevin, Giordano2024Effective}. Secondly, these results may even be generalized to quantum Brownian motion \cite{Fisher1985Quantum}. Thirdly, these results can be experimentally verified using levitated particles \cite{Raynal2023Shortcuts, Gieseler2012Subkelvin, Kamba2023Revealing} and ultracold atoms \cite{Kindermann2017Nonergodic, Sancho2004Diffusion}. However, the Einstein relation will not be sustained for any bounded potential. It has been found that in the time-dependent potentials, some new universal laws can be achieved \cite{Krivolapov2012Transport, Krivolapov2012Universality, Jones2004Rectifying}, thus it is important to explore the possible condition for the failure of the above results, leading to new physics related to non-Gaussian PDF, ergodicity breaking and non-equilibrium dynamics.  

\textit{Acknowledgments}: We thank Prof. Jing-dong Bao and Prof. Hao Ge for their valuable discussion. This work is supported by the Strategic Priority Research Program of the Chinese Academy of Sciences (Grant No. XDB0500000),  and the Innovation Program for Quantum Science and Technology (2021ZD0301200, 2021ZD0301500).

\bibliography{ref.bib}
\end{document}